\documentclass[twocolumn,nopacs,floatfix,amsmath,nofootinbib,amssymb,preprintnumbers,aps]{revtex4}
\usepackage{graphicx,color,dcolumn,booktabs,bm}
\usepackage{txfonts}
\usepackage{slashed,bbm}
\usepackage{epstopdf}
\usepackage{array,makecell,tabularx}
\usepackage[colorlinks,
            citecolor=blue,
            anchorcolor=red,
            menucolor=red,
            linkcolor=red,
            filecolor=red,
            runcolor=red,
            urlcolor=blue,
            frenchlinks=red]{hyperref}

\begin{document}
\title{Analysis of $\Sigma^*$ via isospin selective reaction $K_Lp \to \pi^+\Sigma^0$ }
\author{Dan Guo$^{1}$}\email{guod13@pku.edu.cn}
\author{Jun Shi$^{2,3}$}\email{jun.shi@scnu.edu.cn}
\author{Igor Strakovsky$^{4}$}
\author{Bing-Song Zou$^{5,6,7}$}

\affiliation{
$^1$School of Physics and State Key Laboratory of Nuclear Physics and Technology, Peking University,
Beijing 100871, China\\
$^2$State Key Laboratory of Nuclear Physics and Technology, Institute of Quantum Matter, South China Normal University, Guangzhou 510006, China\\
$^3$Guangdong Basic Research Center of Excellence for Structure and Fundamental Interactions of Matter, Guangdong Provincial Key Laboratory of Nuclear Science, Guangzhou 510006, China\\
$^4$Institute for Nuclear Studies, Department of Physics,
The George Washington University, Washington, DC 20052, USA\\
$^5$Department of Physics, Tsinghua University, Beijing 100084, China\\
$^6$Institute of Theoretical Physics, Chinese Academy of Sciences, Beijing 100190, China\\
$^7$Southern Center for Nuclear-Science Theory (SCNT), Institute of Modern Physics,
Chinese Academy of Sciences, Huizhou 516000, China\\
}
\begin{abstract}
     The isospin-selective reaction $K_Lp \to \pi^+\Sigma^0$ provides a clean probe for investigating $I=1$ $\Sigma^*$ resonances. In this work, we perform an analysis of this reaction using an effective Lagrangian approach for the first time, incorporating the well-established $\Sigma(1189) 1/2^+$, $\Sigma(1385) 3/2^+$, $\Sigma(1670) 3/2^-$, $\Sigma(1775) 5/2^-$ states, while also exploring contributions from other unestablished states. 
     By fitting the available differential cross section and recoil polarization data, adhering to partial-wave phase conventions same as PDG, we find that besides the established resonances, contributions from $\Sigma(1660) 1/2^+$, $\Sigma(1580) 3/2^-$ and a $\Sigma^*(1/2^-)$ improve the description. 
     Notably, a $\Sigma^*(1/2^-)$ resonance with mass around 1.54 GeV, consistent with $\Sigma(1620)1/2^-$, is found to be essential for describing the data in this channel, a stronger indication than found in previous analyses focusing on $\pi\Lambda$ final states. 
     While providing complementary support for $\Sigma(1660) 1/2^+$ and $\Sigma(1580) 3/2^-$, our results highlight the importance of the $\Sigma(1620) 1/2^-$ region in $K_Lp \to \pi^+\Sigma^0$. Future high-precision measurements are needed to solidify these findings and further constrain the $\Sigma^*$ spectrum. 
\end{abstract}

\maketitle

\section{Introduction}

Baryon spectroscopy reveals key aspects of non-perturbative QCD and hadron structure. 
Compared with extensively studied nucleon resonances, the $\Sigma^*$ spectrum remains less understood, although it is vital to explore the flavor symmetry of SU(3) and its breaking.

Many $\Sigma^*$ states listed by the Particle Data Group (PDG)~\cite{ParticleDataGroup:2024cfk} still have large uncertainties in their properties or even have questionable existence, as indicated by their low star ratings. 
The situation is especially ambiguous for the long-controversial state $\Sigma(1620)1/2^-$, whose parameters vary considerably between different experimental analyses and theoretical models. 
The quenched quark models~\cite{Capstick:2000qj, Capstick:1986ter, Glozman:1995fu, Loring:2001kx} predict the mass of $J^P= 1/2^-$ $\Sigma^*$ to be around 1650 MeV, whereas the unquenched quark models~\cite{Helminen:2000jb, Zhang:2004xt, Zou:2010tc, Zou:2007mk} suggest a lower mass around 1400 MeV, as the lowest SU(3) partner of $\Lambda(1405)$ within the baryon nonet. 
Our previous analyses~\cite{Wu:2009tu, Wu:2009nw, Gao:2010hy} of the $K^-p\to\Lambda\pi^+\pi^-$ and $\gamma n\to K^+\Sigma^{-}(1385)$ reactions indicate the mass of $\Sigma^*(1/2^-)$ around 1380 MeV, supporting the prediction of unquenched quark models. 
In contrast, the multichannel analysis of $\overline{K}N$ reactions favors the existence of $\Sigma(1660) 1/2^+$ instead of $\Sigma(1620) 1/2^-$~\cite{ParticleDataGroup:2014cgo}. 
Using the unitary chiral approach, the $\Sigma^*(1/2^-)$ mass is predicted to lie near $\overline{K}N$ threshold, and its contribution is widely investigated in analysis of $\Lambda_c^+\to\eta\pi^+\Lambda$~\cite{Lyu:2024qgc}, $\Xi_c^+\to \Lambda\overline{K}^0\pi^+$~\cite{Li:2025exm} reactions. 
A more comprehensive discussion can be found in the recent review~\cite{Wang:2024jyk}. 
Given the lack of consistent experimental confirmation, the evidence for $\Sigma(1620) 1/2^-$ listed in the PDG remains highly uncertain. 

The $\overline{K}N\to \pi\Lambda$ reaction serves as a prominent channel for studying $\Sigma^*$ resonances, as it proceeds via pure $s$-channel hyperon intermediate states with strangeness $S=-1$ and isospin $I=1$. 
In our recent works~\cite{Gao:2010ve, Gao:2012zh}, we analyzed high-statistics data on $K^-p\to\pi^0\Lambda$ from Crystal Ball collaboration~\cite{Prakhov:2008dc}, along with the data from $K^-n\to\pi^-\Lambda$ reaction~\cite{Morris:1978ia}. 
By simultaneously fitting the differential cross sections and $\Lambda$ polarization data for both $K^-p\to\pi^0\Lambda$ and $K^-n\to\pi^-\Lambda$, we find that the $\Lambda$ polarization observables strongly favor the existence of a $\Sigma(1660) 1/2^+$ state with a mass near 1635 MeV, whereas the $\Sigma(1620) 1/2^-$ is not needed by data. 

Besides the $K^-$ beam scattering experiments, a complementary approach utilizes $K_L$ beams, which, as CP eigenstates, contain a $S=-1$ $\overline{K}^0$ component that interacts with protons to produce hyperons. 
The reaction $K_Lp \to \pi^+\Sigma^0$ is isospin-selective process, as it proceeds only via $\overline{K}^0p$ and isolates the pure isospin $I=1$ amplitude, offering a clean probe for studying $\Sigma^*$ resonances.

Moreover, since different analyses are largely based on the same experimental datasets, it is not entirely appropriate to treat their determinations of resonance parameters as independent or to average them~\cite{ParticleDataGroup:2014cgo}.  In the future, complementary spin-selective scattering data induced by $K_L$ beams will provide valuable insights into hyperon spectroscopy. 

Although less statistic than $K^-$ scattering, past measurements of $K_Lp \to \pi^+\Sigma^0$ provide data on differential cross sections and $\Sigma^0$ recoil polarization, though often with limited precision~\cite{Amaryan:2016ufk, GlueX:2017hgs, KLF:2020gai}. In this work, we perform a phenomenological analysis of this isospin-selective reaction using the effective Lagrangian approach to explore the $\Sigma^*$ spectrum. 
Our model incorporates the $s$-channel $\Sigma$ and its resonances, $t$-channel $K^*$ exchange, and $u$-channel nucleon exchange, with adhering to rigorous partial-wave phase conventions consistent with PDG. 
Fitting to the $K_Lp$ scattering differential cross sections and polarizations, we investigate the contributions of established and focus particularly on the roles of $\Sigma(1660) 1/2^+$ and $\Sigma(1620) 1/2^-$. 

The paper is organized as follows: Sec.~\ref{sec:formalism} outlines the theoretical formalism, detailing the effective Lagrangians and the calculation framework. 
Sec.~\ref{sec:result} presents the results of our fits to the experimental data and discusses their physical implications for the properties of $\Sigma^*$ resonances. Finally, Sec.~\ref{sec:summary} provides a summary of our main conclusions.

\section{Theoretical formalism}\label{sec:formalism}
Given that the mean life $\tau$ of the $K_L$ (51.16 ns) is significantly longer than that of the $K^-$ (12.38 ns) meson, a comparison between low-energy $K_L p$ and $K^- p$ scattering measurements is instructive.
In the precision level of hadron scattering, where CP-violating effects are negligible, the $K_L$ and $K_S$ may be treated as $CP$ eigenstates:
\begin{align}\label{eq:KLprojection}
    K_L&=\frac{1}{\sqrt{2}}(K^0-\overline{K}^0),\nonumber\\
    K_S&=\frac{1}{\sqrt{2}}(K^0+\overline{K}^0).
\end{align}

The $K_L p$ scattering amplitude leading to $\pi \Sigma$ final states is isospin-selective, satisfying
\begin{align}\label{eq_KL2K0b}
    T(K_L p\to\pi^+\Sigma^0) &= -\frac{1}{2}T^1(\overline{K}N\to\pi\Sigma), \\
    T(K_L p\to\pi^0\Sigma^+) &= \frac{1}{2}T^1(\overline{K}N\to\pi\Sigma),
\end{align}
where only the isospin-1 component contributes. In contrast, the $K^- p\to \pi^\pm\Sigma^\mp$ reaction lacks such selectivity \cite{Amaryan:2016ufk},
\begin{align}
    T(K^-p\to\pi^+\Sigma^-)&=\frac{1}{2}T^1(\overline{K}N\to\pi\Sigma)-\frac{1}{\sqrt{6}}T^0(\overline{K}N\to\pi\Sigma),\\
    T(K^-p\to\pi^-\Sigma^+)&=-\frac{1}{2}T^1(\overline{K}N\to\pi\Sigma)-\frac{1}{\sqrt{6}}T^0(\overline{K}N\to\pi\Sigma),\\
    T(K^-p\to\pi^0\Sigma^0)&=\frac{1}{\sqrt{6}}T^0(\overline{K}N\to\pi\Sigma).
\end{align}

\begin{figure}[htbp]
    \centering
\includegraphics[width=0.48\textwidth]{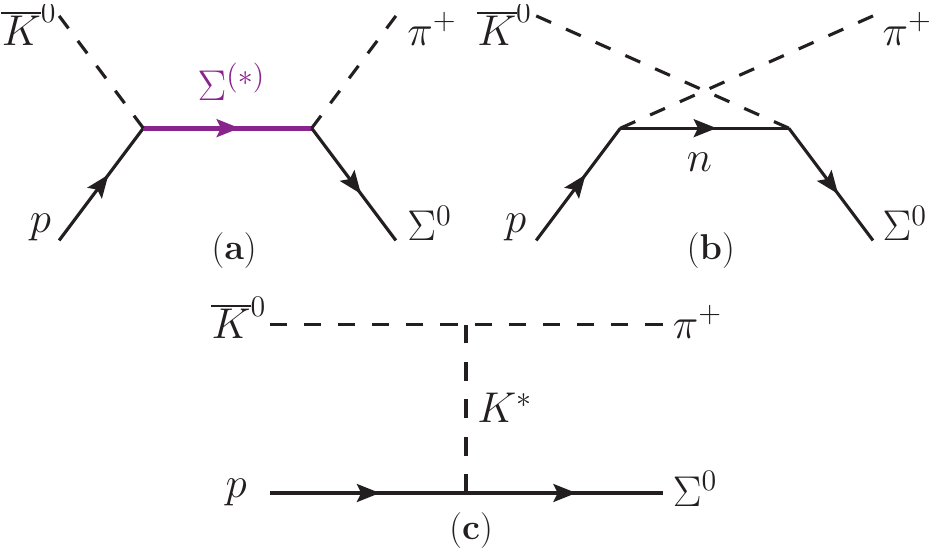}
    \caption{The tree-level Feynman diagrams of the process $\overline{K}^0(k_1) p(p_1) \to \pi^+(k_2) \Sigma^0(p_2)$: (a) $s$-channel $\Sigma$ and its resonances exchange, (b) $u$-channel nucleon exchange, and (c) $t$-channel $K^*$ exchange.}
    \label{Feynman}
\end{figure}

For $\pi^+\Sigma^0$ final state, only the $\overline{K}^0$ component of $K_L$ contributes due to the conservation of strangeness. The tree-level Feynman diagrams for $\overline{K}^0(k_1) p(p_1) \to \pi^+(k_2) \Sigma^0(p_2)$ are depicted in Fig. \ref{Feynman}. The contributions from $s$-channel exchange of $\Sigma$ and its resonances, $u$-channel nucleon exchange and $t$-channel $K^*$ exchange are introduced. 

Considering the investigated energy range and corresponding $\Sigma$ resonance candidates \cite{ParticleDataGroup:2024cfk}, we employ the relevant effective Lagrangians of $s$-channel $\Sigma^{(*)}$ resonances exchange up to $D$-wave with $J^P= 1/2^\pm$, $3/2^\pm$ and $5/2^-$. 

For the $s$-channel ground state $\Sigma(1189)$ and its resonances with $J^P= 1/2^+$, we employ the following effective Lagrangians \cite{Gao:2010ve, Kamano:2013iva}:
\begin{align}
    \mathcal{L}_{KN\Sigma} &=\frac{g_{KN\Sigma}}{M_N+M_\Sigma} \partial_\mu \overline{K}\, \overline{\Sigma} \cdot \tau \gamma^\mu\gamma_5 N + \mathrm{H.c.},\label{eq_KNSigma}\\
    \mathcal{L}_{\pi\Sigma\Sigma} &=i\frac{f_{\pi\Sigma\Sigma}}{m_\pi} \overline{\Sigma}\gamma^{\mu}\gamma_{5}\times \Sigma\cdot \partial_{\mu}\pi+\mathrm{H.c.},
\end{align}
where the underlying isospin structure is given by 
\begin{align}
    &\overline{K} =(K^-,\overline{K}^0), \quad N=\begin{pmatrix}p\\n\end{pmatrix},\\
    &\overline{\Sigma}\cdot\tau = \begin{pmatrix}\overline{\Sigma}^0 & \sqrt{2}\,\overline{\Sigma}^+ \\\sqrt{2}\,\overline{\Sigma}^- & -\overline{\Sigma}^0 \end{pmatrix},\\
    &\pi=\left(\frac{1}{\sqrt{2}}(\pi^+ +\pi^-),  \frac{i}{\sqrt{2}}(\pi^+ -\pi^-), \pi^0\right),\\
    &\Sigma=\left(\frac{1}{\sqrt{2}}(\Sigma^+ +\Sigma^-),  \frac{i}{\sqrt{2}}(\Sigma^+ -\Sigma^-), \Sigma^0\right).
\end{align}
In Refs.~\cite{Gao:2010ve, Kamano:2013iva}, the central coupling constants are estimated from SU(3) flavor symmetry, with $g_{KN\Sigma}=3.58$ and $f_{\pi\Sigma\Sigma}=2(-1+\alpha)f_{\pi NN}$, $f_{\pi NN}=\sqrt{4\pi\times0.08}$, $\alpha=0.635$. 
In our analysis, to account for the potential SU(3) symmetry breaking effects, we introduce a tunable scaling factor between $1/2$ to 2 applied to the central value of the coupling constant $g_{KN\Sigma}f_{\pi\Sigma\Sigma}$.
These constraints allow for a more comprehensive exploration of the parameter space and enhances the robustness of our results.

For the $J^P=1/2^-$ $s$-channel intermediate $\Sigma$ resonance, we employ the effective Lagrangians
\begin{align}
    \mathcal{L}_{KN\Sigma(1/2^-)} &= -ig_{KN\Sigma(1/2^-)} \overline{K}\, \overline{\Sigma}\left(1/2^-\right) \cdot \tau N + \mathrm{H.c.},\\
    \mathcal{L}_{\pi\Sigma\Sigma(1/2^-)} &= g_{\pi\Sigma\Sigma(1/2^-)} \overline{\Sigma}\left(1/2^-\right) \times \Sigma \cdot\pi + \mathrm{H.c.}.
\end{align}
In fitting process, the coupling constant product $g_{KN\Sigma(1/2^-)}\cdot g_{\pi\Sigma\Sigma(1/2^-)}$ is set in the magnitude estimated from its decay width.

For the $J^P=3/2^+$ $s$-channel intermediate $\Sigma^*$ resonance, we adopt the effective Lagrangians
\begin{align}
    \mathcal{L}_{KN\Sigma^*} &= \frac{f_{KN\Sigma^*}}{m_K} \partial_\mu \overline{K}\, \overline{\Sigma}^{*\mu} \cdot \tau N + \mathrm{H.c.},\\
    \mathcal{L}_{\pi\Sigma\Sigma^*} &= i\frac{f_{\pi\Sigma\Sigma^*}}{m_\pi} \partial_\mu \pi\cdot \overline{\Sigma}^{*\mu} \times\Sigma + \mathrm{H.c.}.
\end{align}
The coupling constant $f_{\pi\Sigma\Sigma^*}$ for $\Sigma^*(1385)$ is estimated as 0.68 from its decay width $\Gamma_{\Sigma^*\to\Lambda\pi}\approx$4.2 MeV \cite{ParticleDataGroup:2024cfk}, which is close to the value 0.89 evaluated from SU(3) flavor symmetry \cite{Doring:2010ap}. 
For the $KN\Sigma^*$ coupling, we adopt $f_{KN\Sigma^*}=-3.22$ from the estimation of SU(3) flavor symmetry \cite{Oh:2007jd}. 
A tunable factor ranging from $1/2$ to 2 is multiplied to the central $f_{KN\Sigma^*} f_{\pi\Sigma\Sigma^*}$ constant due to SU(3) symmetry breaking effect. 

For the $J^P=3/2^-$ $s$-channel intermediate $\Sigma$ resonance, we employ the effective Lagrangians
\begin{align}
    \mathcal{L}_{KN\Sigma(3/2^-)} &=\frac{f_{KN\Sigma(3/2^-)}}{m_K} \partial_\mu \overline{K}\, \overline{\Sigma}^\mu\left(3/2^-\right) \cdot\tau\gamma_5N+\mathrm{H.c.},\\
    \mathcal{L}_{\pi\Sigma\Sigma(3/2^-)} &= i\frac{f_{\pi \Sigma\Sigma(3/2^-)}}{m_\pi}\partial_\mu\pi\cdot \overline{\Sigma}^\mu\left(3/2^-\right) \times\gamma_5 \Sigma +\mathrm{H.c.}. 
\end{align}
The central coupling constants $f_{KN\Sigma(3/2^-)}/m_K$=5.2 and $f_{\pi \Sigma\Sigma(3/2^-)}/m_\pi$=15.7 are estimated from the decay widths of $\Sigma(1670)$ $D_{13}$ resonance. Due to the large uncertainties of $\Gamma_{\Sigma(1670)\to KN}$ and $\Gamma_{\Sigma(1670)\to \Sigma\pi}$, we account for possible variations by scaling the product $f_{KN\Sigma(3/2^-)}f_{\pi \Sigma\Sigma(3/2^-)}/(m_\pi m_K)$ with a factor varying from $1/2$ to 2.

For the $J^P=5/2^-$ $s$-channel intermediate $\Sigma$ resonance, we adopt the effective Lagrangians
\begin{align}
    \mathcal{L}_{KN\Sigma(5/2^{-})} &=g_{KN\Sigma(5/2^{-})}\partial_{\mu}\partial_{\nu}\overline{K}\, \overline{\Sigma}^{\mu\nu}\left(5/2^-\right)\cdot\tau N +\mathrm{H.c.},\\
    \mathcal{L}_{\pi\Sigma\Sigma(5/2^{-})} &=ig_{\pi\Sigma\Sigma(5/2^{-})}\partial_{\mu}\partial_{\nu}\pi \cdot \overline{\Sigma}^{\mu\nu}\left(5/2^-\right)\times \Sigma +\mathrm{H.c.}.
\end{align}
The central coupling constants $g_{KN\Sigma(5/2^-)}$=7.7 and $g_{\pi \Sigma\Sigma(5/2^-)}= 2.4$ are estimated from the decay widths of $\Sigma(1775)$ $D_{15}$ resonance. Since the large uncertainties of $\Gamma_{\Sigma(1775)\to KN}$ and $\Gamma_{\Sigma(1775)\to \Sigma\pi}$, we also constrain $g_{KN\Sigma(5/2^-)}g_{\pi \Sigma\Sigma(5/2^-)}$ multiplying a factor varying from $1/2$ to 2. 

The interaction of higher-spin fields is known to suffer from the Johnson-Sudarshan-Velo-Zwanziger (JS-VZ) problem~\cite{Johnson:1960vt, Velo:1969bt}, wherein unphysical lower-spin components may appear. This pathology can be resolved via a gauge-invariant field redefinition~\cite{Pascalutsa:1998pw, Pascalutsa:1999zz, Pascalutsa:2000kd}, which is applied in descriptions of pion photoproduction~\cite{Fernandez-Ramirez:2005nuq} and $\pi N$ scattering~\cite{Alarcon:2012kn, Shklyar:2004dy}. However, there don't demonstrate significant numerical advantage over the conventional coupling scheme~\cite{Shklyar:2004dy}. In fact, for tree-level phenomenological calculations, the two formalisms yield identical results~\cite{Cao:2024nxm, Descotes-Genon:2019dbw}. Therefore, consistent with common practice in scattering analyses~\cite{Ronchen:2012eg, Wang:2025gfv}, we employ the conventional couplings throughout this study. 

For $u$-channel nucleon exchange depicted in Fig. \ref{Feynman} (b), we adopt the effective Lagrangians from Eq.~\eqref{eq_KNSigma} together with 
\begin{equation}
    \mathcal{L}_{\pi NN}=\frac{g_{\pi NN}}{2M_N}\bar{N}\gamma^\mu\gamma_5\partial_\mu\pi\cdot\tau N,
\end{equation}
where $g_{\pi NN}$=13.26 is determined from SU(3) flavor symmetry \cite{Gao:2010ve}. 

For the $t$-channel $K^*$ exchange shown as in Fig. \ref{Feynman} (c), we adopt the effective Lagrangians 
\begin{align}
    \mathcal{L}_{K^*K\pi}&=ig_{K^*K\pi}\overline{K}_\mu^*\left(\pi\cdot\tau\partial^\mu K-\partial^\mu\pi\cdot\tau K\right),\\
    \mathcal{L}_{K^*N\Sigma}&=-g_{K^*N\Sigma}\overline{\Sigma}\cdot\tau \left(\gamma_\mu \overline{K}^{*\mu}-\frac{\kappa_{K^*N\Sigma}}{2M_N}\sigma_{\mu\nu}\partial^\nu \overline{K}^{*\mu}\right)N + \mathrm{H.c.}.
\end{align}
The coupling constant $g_{K^*K\pi}=-3.23$ is determined from the decay width of $K^*\to K\pi$, where the negative sign follows from SU(3) relations \cite{Mueller-Groeling:1990uxr}. For the $K^*N\Sigma$ vertex, we consider two sets of coupling parameters from Refs.~\cite{Stoks:1999bz, Oh:2004wp,Shi:2014vha}
\begin{align}
    g_{K^*N\Sigma}&=-2.46,\quad\kappa_{K^*N\Sigma}=-0.47(NSC97a),\\
    g_{K^*N\Sigma}&=-3.52,\quad\kappa_{K^*N\Sigma}=-1.14(NSC97f).
\end{align}
Here, NSC97a and NSC97f refer to two versions of the Nijmegen soft-core potential model, which provide SU(3)-based predictions for baryon-baryon couplings.
Taking into account of theoretical  uncertainty, we constrain the couplings $g_{K^*N\Sigma}$ within the range from -7.0 to -1.2, and $\kappa_{K^*N\Sigma}$ within the range from -2.3 to -0.2. 

At each vertex, we introduce a form factor to account for off-shell effects:
\begin{equation}
    F_B\left(q_{ex}^2,M_{ex}\right)=\frac{\Lambda^4}{\Lambda^4+(q_{ex}^2-M_{ex}^2)^2},
\end{equation}
where $q_{ex}$ and $M_{ex}$ denote the four-momentum and mass of the exchange particle, respectively. In our analysis, the cutoff $\Lambda$ is constrained to the range from 0.5 to 2 GeV for all diagrams. 

The propagator for the $t$-channel $K^*$ intermediate state is given by
\begin{equation}
    G_{K^*}(p)=\frac{-g^{\mu\nu}+p^\mu p^\nu/m_{K^*}^2}{p^2-m_{K^*}^2}.
\end{equation}
Similarly, the propagator for the $u$-channel nucleon exchange takes the form
\begin{equation}
    G_B(q)=\frac{\slashed{q}+M_N}{q^2-M_N^2}.
\end{equation}
The propagator for the $s$-channel spin-$1/2$ $\Sigma$ resonance exchange reads as
\begin{equation}
    G_R^{1/2}(q)=\frac{\slashed{q}+M}{q^2-M^2+iM\Gamma}.
\end{equation}
For spin-$3/2$ $\Sigma$ resonance exchange, the propagator takes the form
\begin{align}
    G_R^{3/2}(q)=&\frac{\slashed{q}+M}{q^2-M^2+iM\Gamma}\left(-g^{\mu\nu}+\frac{\gamma^\mu\gamma^\nu}{3} \right. \nonumber\\ 
    & +\frac{\gamma^\mu q^\nu-\gamma^\nu q^\mu}{3M}+\frac{2q^\mu q^\nu}{3M^2}\left.\right).
\end{align}
Similarly, for the spin-$5/2$ $\Sigma$ resonance exchange, the propagator is given by
\begin{equation}
    G_R^{5/2}(q)=\frac{\slashed{q}+M}{q^2-M^2+iM\Gamma}S_{\alpha\beta\mu\nu}(q,M),
\end{equation}
where 
\begin{align}
    S_{\alpha\beta\mu\nu}(q,M)=&\frac{1}{2}(\bar{g}_{\alpha\mu}\bar{g}_{\beta\nu}+\bar{g}_{\alpha\nu}\bar{g}_{\beta\mu})-\frac{1}{5}\bar{g}_{\alpha\beta}\bar{g}_{\mu\nu}-\frac{1}{10}(\bar{\gamma}_\alpha\bar{\gamma}_\mu\bar{g}_{\beta\nu} \nonumber\\
    &+\bar{\gamma}_\alpha\bar{\gamma}_\nu\bar{g}_{\beta\mu}+\bar{\gamma}_\beta\bar{\gamma}_\mu\bar{g}_{\alpha\nu}+\bar{\gamma}_\beta\bar{\gamma}_\nu\bar{g}_{\alpha\mu}),
\end{align}
with
\begin{align}
    &\bar{g}_{\mu\nu}=g_{\mu\nu}-\frac{q_{\mu}q_{\nu}}{M^{2}},\\
    &\bar{\gamma}_{\mu}=\gamma_{\mu}-\frac{q_{\mu}}{M^{2}}\slashed{q}.
\end{align}

Using the Lagrangians and propagators introduced above, one can construct the scattering amplitude for the process $\overline{K}^0(k_1) +p(p_1)\to\pi^+(k_2) +\Sigma^0(p_2)$, denoted as $\mathcal{M}_{\overline{K}^0p\to\pi^+\Sigma^0}$, 
\begin{equation}
    \mathcal{M}_{\overline{K}^0p\to\pi^+\Sigma^0}^{r_2,r_1}=\bar{u}_{r_2}(p_2)\mathcal{A}u_{r_1}(p_1)=\bar{u}_{r_2}(p_2)\left(\sum_i\mathcal{A}_i\right)u_{r_1}(p_1),
\end{equation}
where $r_1$ and $r_2$ denote the initial nucleon and final $\Sigma^0$ polarizations, respectively, and $\mathcal{A}_i$ represents the non-spinor part of the amplitude corresponding to the $i$th diagram.

Then the differential cross section for the process $K_L p\to\pi^+ \Sigma^0$ is expressed as 
\begin{equation}
    \frac{\textmd{d}\sigma_{K_Lp\to\pi^+\Sigma^0}}{\textmd{d}\Omega}=\frac{\textmd{d}\sigma_{K_Lp\to\pi^+\Sigma^0}}{2\pi \textmd{d} \cos(\theta)}=\frac{1}{2}\frac{1}{64\pi^2s}\frac{|\vec{k}_2|}{|\vec{k}_1|}\overline{|\mathcal{M}_{\overline{K}^0p\to\pi^+\Sigma^0}|}^2,
\end{equation}
where the factor $\frac{1}{2}$ arises from the projection $K_L$ to $\overline{K}^0$, as given in Eq.~\eqref{eq:KLprojection}. The spin-averaged squared amplitude is expressed as
\begin{align}
    \overline{|\mathcal{M}_{\overline{K}^0p\to\pi^+\Sigma^0}|}^2 &=\frac{1}{2}\sum_{r_1,r_2}\mathcal{M}_{\overline{K}^0p\to\pi^+\Sigma^0}^{r_2,r_1} \mathcal{M}_{\overline{K}^0p\to\pi^+\Sigma^0}^{\dagger r_2,r_1} \nonumber\\
    &=\frac{1}{2}\mathrm{Tr}\left[(\slashed{p}_2+m_\Sigma)\mathcal{A} (\slashed{p}_1+m_N) \gamma^0\mathcal{A}^+\gamma^0 \right].
\end{align}

In addition to the differential cross section, we also calculate the recoil polarization of the $\Sigma^0$ baryon When both the incident beam and the target are unpolarized. The recoil polarization is defined as Refs~\cite{Amaryan:2016ufk, Shi:2014vha, Penner:2002ma}
\begin{equation}
    P_{\Sigma}=-\frac{2\mathrm{Im}{\left( \mathcal{M}_{\overline{K}^0p\to\pi^+\Sigma^0}^{1/2,1/2} \mathcal{M}_{\overline{K}^0p\to\pi^+\Sigma^0}^{*-1/2,1/2}\right)} } {\overline{|\mathcal{M}_{\overline{K}^0p\to\pi^+\Sigma^0}|}^2},
\end{equation}
which describes the asymmetry in the spin distribution of the recoiling $\Sigma^0$ along the direction $\hat{\mathbf{v}}=\hat{\bf{k_1}}\times\hat{\bf{k_2}}$, normal to the reaction plane.

\section{Result and discussion}\label{sec:result}

\setlength{\extrarowheight}{5pt} 
\begin{table}[htbp]
    \centering
    \caption{The fixed masses, widths and signs of the imaginary parts of amplitudes at resonance points of $\Sigma(1189) 1/2^+$, $\Sigma(1385) 3/2^+$, $\Sigma(1670) 3/2^-$, and $\Sigma(1775) 5/2^-$. }
    \label{tab:fixed_para}
    \setlength{\tabcolsep}{10pt} 
    \begin{tabular}{lccc}
    \toprule[1.50pt]
    \toprule[0.50pt]
               & $m$ [MeV] &     $\Gamma$ [MeV] &     Signs \\ 
    \hline
    $\Sigma(1189) 1/2^+$ & 1192 & 0 & N/A \\
    $\Sigma(1385) 3/2^+$ & 1384 & 36 & $\downarrow$ \\
    $\Sigma(1670) 3/2^-$ & 1675 & 70 & $\uparrow$ \\
    $\Sigma(1775) 5/2^-$ & 1775 & 120 & $\uparrow$ \\
    \bottomrule[0.50pt]
    \bottomrule[1.50pt]
    \end{tabular}
\end{table}
\setlength{\extrarowheight}{0pt}

Using the differential cross section and polarization formalism introduced above, and allowing the couplings in each diagram to vary by a factor ranging from $1/2$ to 2, we achieve a good description of the differential cross section and recoil $\Sigma^0$ polarization data.

As a background contribution, we include the $u$-channel nucleon exchange and the $t$-channel $K^*$ exchange, which account for the underlying dynamics of the process, in contrast to purely phenomenological polynomial parameterizations that lack a direct physical dynamics. 
In addition, we incorporate the well-established four-star $\Sigma$ resonances below 1.8 GeV: $\Sigma(1189) 1/2^+$, $\Sigma(1385) 3/2^+$, $\Sigma(1670) 3/2^-$, and $\Sigma(1775) 5/2^-$. These states are always included in our analysis, and their necessity is further supported by attempts in which switching off any of these resonances results in a significantly worse description of the data.
Furthermore, we examine the contributions of the three-star $\Sigma(1660) 1/2^+$ and $\Sigma(1750) 1/2^-$, as well as the one-star $\Sigma(1580) 3/2^-$ and $\Sigma(1620) 1/2^-$, to confirm their potential existence and evaluate their role in improving the fit to experimental data. 

The fixed masses, widths and sign conventions of four-star resonances used in the fitting procedure are listed in Table~\ref{tab:fixed_para}. The sign conventions for the resonance couplings in the $\overline{K}N\to\pi\Sigma$ are aligned with partial-wave analyses of experimental data and SU(3) flavor assignments~\cite{LeviSetti:1969zz}, as outlined in the PDG review ``\textit{$\Lambda$ and $\Sigma$ Resonances}"~\cite{ParticleDataGroup:2024cfk}. In general, the signs of $\Sigma(1385)$ and $\Lambda(1405)$ are fixed by convention, while those of other resonances are determined relative to them.
When including only the background contributions and the established four-star $\Sigma^*$ resonances, the fit yields a poor result with $\chi^2/\text{D.o.F} = 2.789$ using 12 parameters, indicating the necessity of additional resonance contributions to better describe the data. 

\begin{figure*}[htbp]
    \centering
\includegraphics[width=\textwidth]{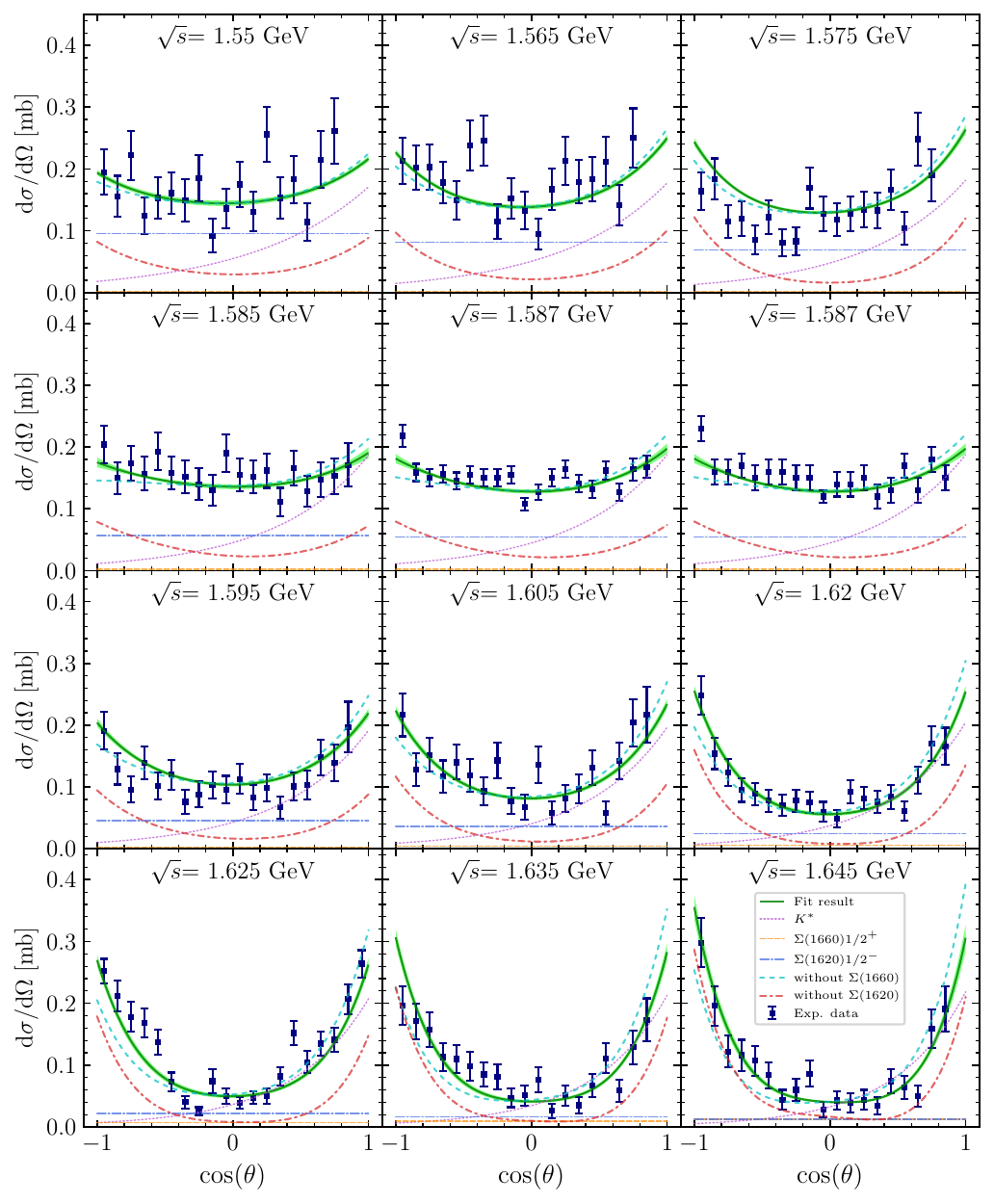}
    \caption{The fitted differential cross sections of $K_Lp \to \pi^+\Sigma^0$ reaction, shown with 1-$\sigma$ uncertainty bands. Contributions from the $t$-channel $K^*$ exchange, and the $s$-channel $\Sigma(1660) 1/2^+$ and $\Sigma(1620) 1/2^-$ resonances are shown. Results excluding either the $\Sigma(1660) 1/2^+$ or $\Sigma(1620) 1/2^-$ contribution are also plotted for comparison. 
    The square marks represent the experimental data with errors: the first set data at $\sqrt{s}$= 1.587 GeV from Ref.~\cite{Cho:1975dv}, the second set data at 1.587 GeV from Ref.~\cite{Engler:1978rr}, the data at 1.625 GeV from Ref.~\cite{Burkhardt:1975sd}, and other data from Ref.~\cite{Bologna-Edinburgh-Glasgow-Pisa-Rutherford:1977cyz}. }
    \label{fig:dcs}
\end{figure*}

\begin{figure*}[htbp]
    \centering
\includegraphics[width=\textwidth]{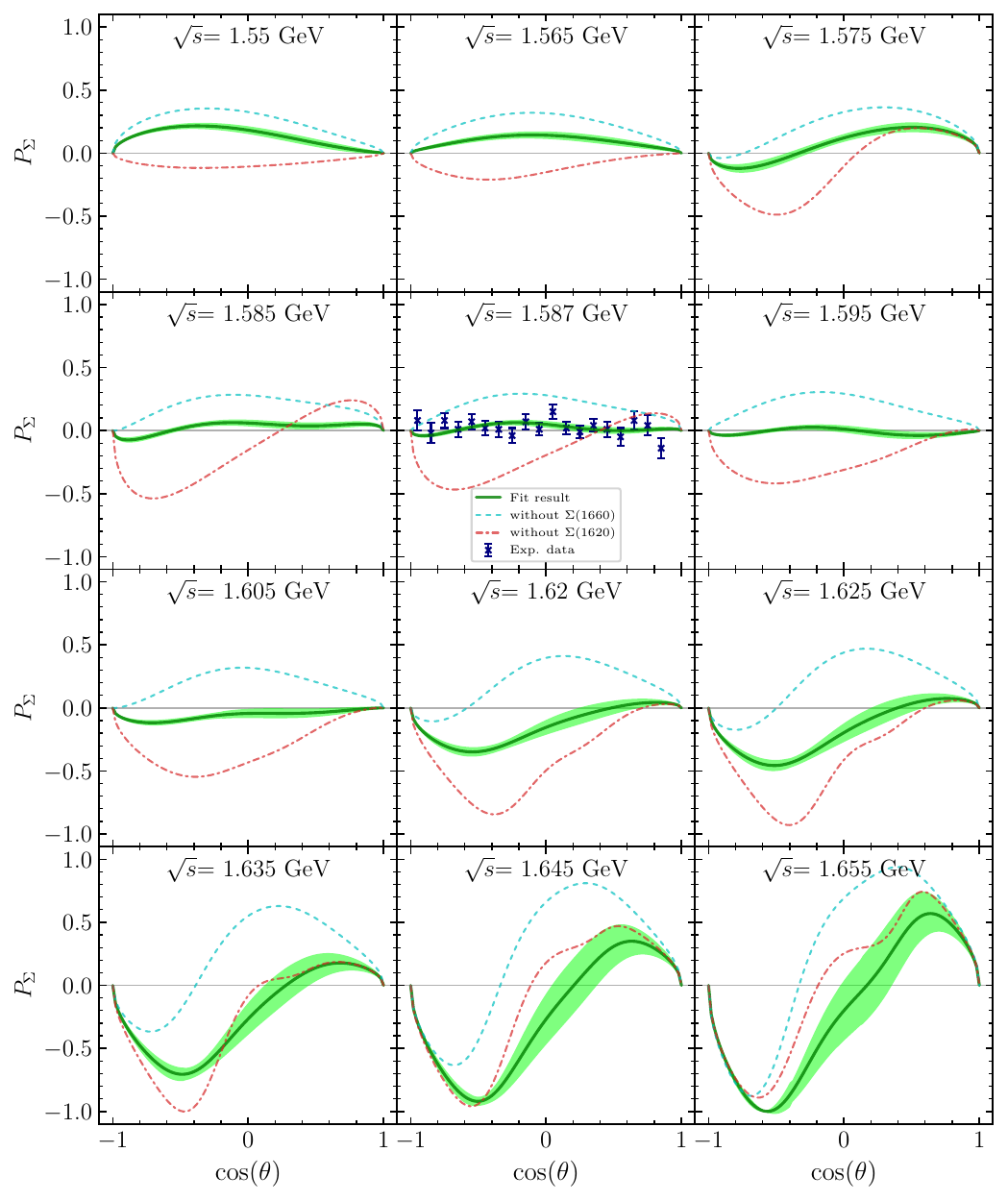}
    \caption{The recoil polarizations of $\Sigma^0$ hyperon in $K_Lp \to \pi^+\Sigma^0$ reaction, shown with 1-$\sigma$ uncertainty bands. Results excluding either the $\Sigma(1660) 1/2^+$ or $\Sigma(1620) 1/2^-$ contribution are also plotted for comparison. 
    The experimental data at 1.587 GeV from Ref.~\cite{Engler:1978rr} are included in the fitting procedure. }
    \label{fig:polar}
\end{figure*}

The coupling parameters in the above diagrams remain uncertain and are fitted in the analysis, with their ranges constrained by PDG estimates and model predictions, as discussed in Sec.~\ref{sec:formalism}.
Building upon the fundamental contributions from $t$-channel and $u$-channel exchanges, as well as $s$-channel exchanges of the well-established four-star resonances, we further explore various combinations by including possible contributions from $\Sigma(1660) 1/2^+$,$\Sigma(1620) 1/2^-$, $\Sigma(1750) 1/2^-$ and $\Sigma(1580) 3/2^-$. 
The resulting optimal fit to the available differential cross section data~\cite{Cho:1975dv, Engler:1978rr, Burkhardt:1975sd, Bologna-Edinburgh-Glasgow-Pisa-Rutherford:1977cyz} and $\Sigma^0$ recoil polarization data~\cite{Engler:1978rr} is shown in Fig.~\ref{fig:dcs} and Fig.~\ref{fig:polar}. 
With the inclusion of the additional resonances $\Sigma(1660) 1/2^+$, $\Sigma(1620) 1/2^-$, and $\Sigma(1580) 3/2^-$, the model comprises a total of 9 diagrams and 22 tunable parameters, including form factor cutoffs. 
The optimal fit yields a reduced chi-squared of $\chi^2/\text{D.o.F} = 1.606$ ($\chi^2 = 358.104$) for a total of 245 data points. The fitted parameters, along with their corresponding estimates from decay widths or SU(3) symmetry predictions, are summarized in Table~\ref{tab:fitted_para}. 
The parameter uncertainties obtained from the MINUIT fit, including the full covariance matrix from the Hesse method, are propagated to the observables. The resulting 1-$\sigma$ confidence bands are shown in Fig.~\ref{fig:dcs} and Fig.~\ref{fig:polar}. The error bands are narrow for the differential cross sections, indicating strong constraints from the data, while those for recoil polarization are visibly broader, as the fit is primarily driven by the differential cross section measurements.

\renewcommand{\arraystretch}{1.5}  
\begin{table*}[tbph]
    \centering
    \caption{The optimally fitted parameters, and three alternative fitting scenarios for comparison. Bracketed ranges indicate estimates based on SU(3) symmetry relations or PDG-reported values. A dash (---) denotes parameters not included in the fit.}
    \label{tab:fitted_para}
    \setlength{\tabcolsep}{8pt} 
    \begin{tabular}{l|llllll}
    \toprule[1.50pt]
    \toprule[0.50pt]
    Resonances & Parameters & Optimal Fit & Fit I & Fit II & Fit III & Estimates \\ \hline
    $K^*$    & $g_{K^*N\Sigma}$      & -7.0$\pm$0.5   & -7.0  & -2.7& -7.0$\pm$1.2  & [-7.0, -1.2]\\
             & $\kappa_{K^*N\Sigma}$ & -1.6$\pm$0.2   & -2.3  & -2.3& -1.6$\pm$0.8  & [-2.3, -0.2]\\
             & $\Lambda$   & 0.97$\pm$0.01 & 1.01 & 2.0  & 0.97$\pm$0.09  & [0.5, 2.0] \\ \hline
    $N$      & $\Lambda$   & 1.42$\pm$0.04 & 1.93 & 1.47 & 1.4$\pm$0.4    & [0.5, 2.0] \\ \hline
    $\Sigma(1189) 1/2^+$   & $g_{KN\Sigma}f_{\pi\Sigma\Sigma}$ & -1.50$\pm$3.0 & -3.39& -1.35 & -1.50$\pm$3.0 & [-5.4, -1.3] \\
                           & $\Lambda$    & 0.5$\pm$1.1   & 0.6 & 0.5 & 0.5$\pm$1.3   & [0.5, 2.0] \\ \hline
    $\Sigma(1385) 3/2^+$  & $f_{KN\Sigma^*} f_{\pi\Sigma\Sigma^*}$  & -1.34$\pm$4.0 & -4.49 & -5.7 & -1.34$\pm$4.0 & [-5.7, -1.1] \\
                          & $\Lambda$     & 0.50$\pm$0.18 & 0.5 & 0.6 & 0.5$\pm$1.3   & [0.5, 2.0] \\ \hline
    $\Sigma(1670) 3/2^-$  & $\sqrt{\Gamma_{\overline{K}N}\Gamma_{\pi\Sigma} }/\Gamma_{tot}$  & +0.26$\pm$0.04 & +0.27& +0.29 & +0.24$\pm$0.06 & [0.09, 0.38] \\
                          & $\Lambda$     & 0.72$\pm$0.07 & 0.61 & 0.62 & 0.76$\pm$0.17 & [0.5, 2.0] \\ \hline
    $\Sigma(1775) 5/2^-$  & $\sqrt{\Gamma_{\overline{K}N}\Gamma_{\pi\Sigma} }/\Gamma_{tot}$  & +0.24$\pm$0.04 & +0.24& +0.24 & +0.24$\pm$0.12 & [0.06, 0.24] \\
                          & $\Lambda$     & 2.0$\pm$1.4   & 2.0 & 1.1 & 2.0$\pm$1.5   & [0.5, 2.0] \\ \hline
    $\Sigma(1580) 3/2^-$  & $\sqrt{\Gamma_{\overline{K}N}\Gamma_{\pi\Sigma} }/\Gamma_{tot}$  & +0.032$\pm$0.005 & +0.034& --- & +0.031$\pm$0.005 & [-0.4, 0.4] \\
                          & $\Lambda$     & 0.50$\pm$0.09 & 0.5 & --- & 0.50$\pm$0.11 & [0.5, 2.0] \\ \hline
    $\Sigma(1660) 1/2^+$  & $M$ [GeV] & 1.696$\pm$0.010 & 1.75   & --- & 1.673$\pm$0.027 & [1.40, 1.75] \\
                          & $\Gamma$ [GeV] & 0.108$\pm$0.021 & 0.073& --- & 0.10$\pm$0.04 & [0.01, 0.40] \\
                          & $\sqrt{\Gamma_{\overline{K}N}\Gamma_{\pi\Sigma} }/\Gamma_{tot}$  & -0.112$\pm$0.006 & -0.121& --- & -0.086$\pm$0.048 & [-0.48, 0.48] \\
                          & $\Lambda$     & 2.0$\pm$0.8   & 2.0 & --- & 2.0$\pm$1.2   & [0.5, 2.0] \\ \hline
    $\Sigma(1620) 1/2^-$  & $M$ [GeV] & 1.541$\pm$0.003 & 1.4(Fixed) & 1.545 & 1.542$\pm$0.007 & [1.35, 1.65] \\
                          & $\Gamma$ [GeV] & 0.129$\pm$0.002 & 0.4   & 0.10 & 0.16$\pm$0.05 & [0.01, 0.40] \\
                          & $\sqrt{\Gamma_{\overline{K}N}\Gamma_{\pi\Sigma} }/\Gamma_{tot}$  & -0.633$\pm$0.009 & -2.32& -0.45 & -0.779$\pm$0.259 & [-3.2, 3.2] \\
                          & $\Lambda$     & 0.89$\pm$0.04 & 1.07 & 0.59 & 0.72$\pm$0.19 & [0.5, 2.0] \\ \hline
    $\Sigma(1750) 1/2^-$  & $\sqrt{\Gamma_{\overline{K}N}\Gamma_{\pi\Sigma} }/\Gamma_{tot}$  & --- & --- & --- & +0.093$\pm$0.187 & [-1.2, 1.2] \\
                          & $\Lambda$     & ---   & --- & --- & 1.9$\pm$0.8   & [0.5, 2.0] \\ \hline
                    & \text{D.o.F}           & 223    & 224    & 229   & 221     & \\
                    & $\chi^2/\text{D.o.F}$  & 1.606  & 1.707  & 1.774 & 1.619   & \\
    \bottomrule[0.50pt]
    \bottomrule[1.50pt]
    \end{tabular}
\end{table*}
\setlength{\extrarowheight}{0pt}

As shown in Fig.~\ref{fig:dcs}, the model provides a reasonable global description of the $K_Lp \to \pi^+\Sigma^0$ differential cross sections. 
However, the limited quality of the historical data—characterized by large uncertainties and inconsistencies between adjacent c.m. energy points—prevents a more precise constraint on the model parameters. 
We also display the sizable single-diagram contributions from the $t$-channel $K^*$ exchange and the $s$-channel $\Sigma(1660) 1/2^+$, $\Sigma(1620) 1/2^-$ resonances, based on the optimal fit. 
It is clear that the $t$-channel $K^*$ exchange dominates the overall forward-angle enhancement, while the $s$-channel contributions exhibit no pronounced angular dependence. 
In Fig.~\ref{fig:polar}, besides the 1.587 GeV data included in the fit, predictions of the recoil polarization are also provided for other energy points. 
Here, we do not present the single-diagram contributions to the recoil polarization, as the polarization arises essentially from the interference between different reaction mechanisms (i.e., different diagrams) in spin space.
Consequently, individual single-diagram processes do not generate nonzero recoil polarization on their own. 
By selectively excluding each diagram, we find that the pronounced forward- and backward-angle structures observed at higher energies are primarily attributed to the contribution of the $\Sigma(1670) 3/2^-$ resonance. 

From the optimized parameters in Table~\ref{tab:fitted_para}, one observes that the signs of the couplings $\sqrt{\Gamma_{\overline{K}N}\Gamma_{\pi\Sigma} }/\Gamma_{tot}$ for the four-star resonances are consistent with the signs listed in Table~\ref{tab:fixed_para} denoting the signs of the imaginary parts of $\overline{K}N\to\pi\Sigma$ amplitude, with the sign of $\Sigma(1385)3/2^+$ set by convention. 
By adopting appropriate phase conventions for the interaction Lagrangians, as validated in Refs.~\cite{Gao:2010ve, Gao:2012zh, Shi:2014vha}, the overall signs of the coupling products remain consistent with the amplitude phases of the $\overline{K}N \to \pi\Sigma$ channel, while exhibiting overall opposite signs for the $\overline{K}N \to \pi\Lambda$ channel. 
Since the available data lie far from the $\sqrt{s}$ region where the $\Sigma(1189) 1/2^+$ and $\Sigma(1385) 3/2^+$ resonances dominate, the current analysis cannot tightly constrain their couplings, resulting in large uncertainties. 

For the unestablished $s$-channel resonances $\Sigma(1660) 1/2^+$, $\Sigma(1620) 1/2^-$ and $\Sigma(1580) 3/2^-$, the fitting coupling products are constrained within broad unsigned ranges. 
Including the indispensable contribution from the $\Sigma(1660) 1/2^+$ resonance yields the fitted mass of around 1696 MeV, the width around 108 MeV consistent with our previous results in the $\overline{K}N \to \pi\Lambda$ channel~\cite{Gao:2010ve, Gao:2012zh}. The fitted coupling, $\sqrt{\Gamma_{\overline{K}N}\Gamma_{\pi\Sigma} }/\Gamma_{tot}$, is also compatible with PDG range, especially for the negative sign. 
The present fit in the $\overline{K}N \to \pi\Sigma$ channel provides a further complementary support for the existence of the three-star $\Sigma(1660) 1/2^+$ resonance. 
In Fig.~\ref{fig:dcs} and Fig.~\ref{fig:polar}, we also present the results excluding the $\Sigma(1660) 1/2^+$ contributions for comparison. 
Although the $\Sigma(1660) 1/2^+$ provides only a small single-diagram contribution, 
its removal leads to noticeable changes in both the differential cross sections and polarizations, highlighting the significant role of interference effects. 

Regarding the long-debated $\Sigma(1620) 1/2^-$, we constrain its mass varying from 1.35 GeV to 1.65 GeV during the fitting procedure. Unlike our previous work in the $\overline{K}N \to \pi\Lambda$ channel~\cite{Gao:2010ve, Gao:2012zh}, here the contribution of $\Sigma(1620) 1/2^-$ is essential in the $\overline{K}N \to \pi\Sigma$ channel. 
We obtain the fitted mass of 1541 MeV and width of 129 MeV, which is compatible with the mass of 1501 MeV and width of 171 MeV reported in Ref.~\cite{Zhang:2013sva}. Additionally, it is in agreement with the mass of 1551 MeV and width of 188 MeV from Model B of Ref.~\cite{Kamano:2015hxa}. 
While the fitted coupling $\sqrt{\Gamma_{\overline{K}N}\Gamma_{\pi\Sigma} }/\Gamma_{tot}$ exhibit tiny uncertainty, it is estimated by the experimental total width carrying apparent uncertainty. 
The negative sign of coupling is more plausible, and should be further examined with future high-precision measurements. 
The single-diagram contribution of $\Sigma(1620) 1/2^-$ is more prominent at lower energies. Comparing the results with and without the inclusion of $\Sigma(1620) 1/2^-$, as shown in Fig.~\ref{fig:dcs} and Fig.~\ref{fig:polar}, indicates its necessity for an accurate description of the data, as well as the increasingly significant interference effects at higher energies. 

We also examined the scenario where the mass of $\Sigma(1620) 1/2^-$ is fixed at 1.4 GeV, labeled as fit I in Table~\ref{tab:fitted_para} (due to the poor quality of the fit, uncertainties on the parameters are not provided). This scenario leads to a larger $\chi^2/\mathrm{D.o.F}=1.707$, and results in unphysical values for the mass and width of $\Sigma(1660) 1/2^+$. 
Furthermore, when only the $\Sigma(1620) 1/2^-$ resonance is included—excluding both $\Sigma(1660) 1/2^+$ and $\Sigma(1580) 3/2^-$—the fit improves notably. This scenario, labeled as fit III in Table~\ref{tab:fitted_para}, yields $\chi^2/\mathrm{D.o.F} = 1.774$ with 16 adjustable parameters, a clear improvement over the baseline value of 2.789. 
At least, the current $K_LP\to\pi^+\Sigma^0$ scattering data support the necessity of a $\Sigma^*(1/2^-)$ resonance with mass around 1.54 GeV. 

The inclusion of the one-star $\Sigma(1580) 3/2^-$ also improves the fit, with the fixed mass 1.58 GeV and width 0.1 GeV~\cite{ParticleDataGroup:2024cfk}. 
The resulted positive coupling, $\sqrt{\Gamma_{\overline{K}N}\Gamma_{\pi\Sigma}}/\Gamma_{\text{tot}} = +0.032 \pm 0.005$, is in good agreement with the sole value reported by the PDG.

\begin{figure}
    \centering
    \includegraphics[width=0.48\textwidth]{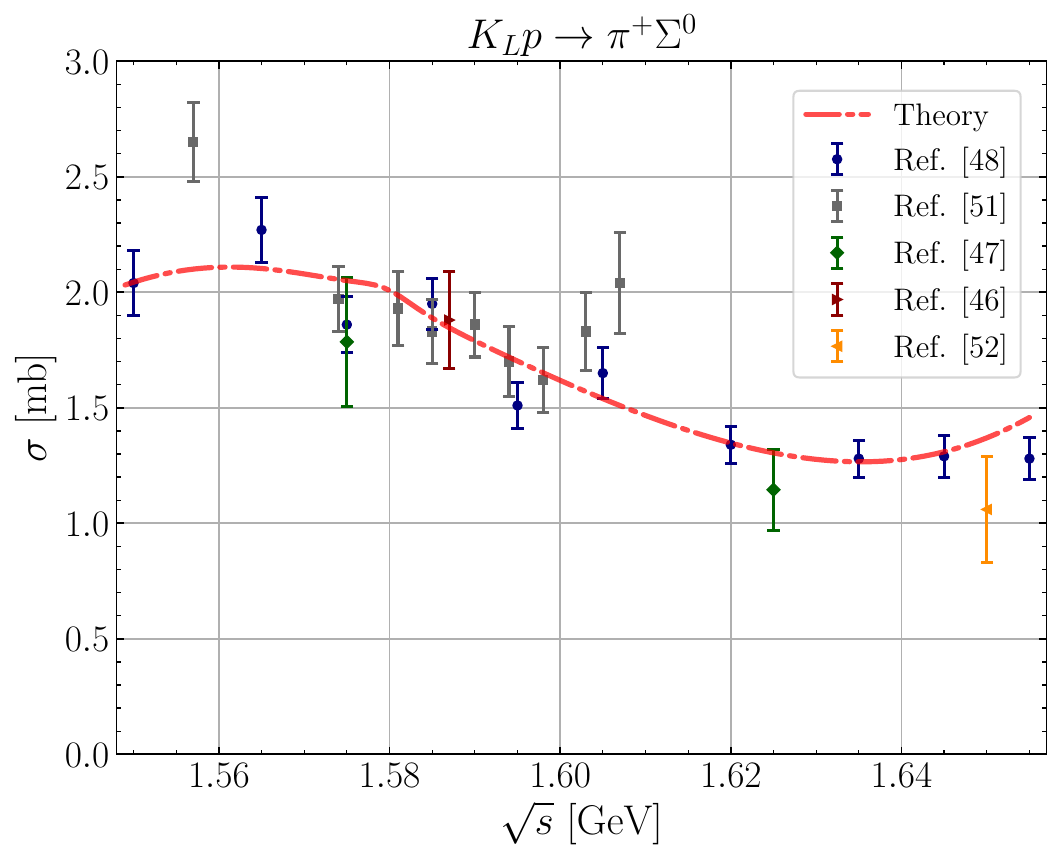}
    \caption{The comparison of predicted $K_LP\to\pi^+\Sigma^0$ total cross section with experimental data~\cite{Bologna-Edinburgh-Glasgow-Pisa-Rutherford:1977cyz, Engler:1976cn, Burkhardt:1975sd, Engler:1978rr, Yamartino:1974sm}. }
    \label{fig:tcs}
\end{figure}

The inclusion of $\Sigma(1750) 1/2^-$, labeled as fit II in Table~\ref{tab:fitted_para}, leads to a marginal improvement in the total $\chi^2 = 357.888$, but results in a slightly worse reduced chi-squared of $\chi^2/\mathrm{D.o.F}=1.619$ compared to the optimal fit. 
This indicates that the current $K_Lp$ scattering data do not strongly support the necessity of a $\Sigma(1750) 1/2^-$ contribution. 

Moreover, based on the globally consistent fit to the differential cross sections and polarization data, we also present a comparison between the theoretical prediction and the experimental measurements of the total cross section~\cite{Bologna-Edinburgh-Glasgow-Pisa-Rutherford:1977cyz, Engler:1976cn, Burkhardt:1975sd, Engler:1978rr,  Yamartino:1974sm}, as shown in Fig.~\ref{fig:tcs}. 
The good agreement between the theoretical prediction and the experimental total cross section supports the validity of our analysis and indicates that the adopted scenario captures the essential features of the reaction dynamics. 

Due to the limitations of the current low-quality data, significant uncertainties of some parameters remain.
Future measurements with broader energy coverage and higher precision, combined with multi-channel analyses—particularly of the pure isospin-1 channels $K_LP\to\pi^+\Sigma^0$ and $K_LP\to\pi^+\Lambda$ will be crucial for gaining deeper insights into the properties and interaction mechanisms of the $\Sigma^*$ resonances. 
In particular, the polarization data provides valuable insight into the interference between different reaction mechanisms. 
Such efforts will be especially important for clarifying the existence and characteristics of the $\Sigma(1620) 1/2^-$ state. 

\section{Summary}\label{sec:summary}

The $CP$ eigenstate $K_L$, as a quantum superposition of the flavor eigenstates $K^0$ and $\overline{K}^0$, serves as a more effective hadronic probe than electromagnetic probes $\gamma^{(*)}$, which couple less directly to hadrons. The $S=-1$ $\overline{K}^0$ component contributes significantly more—approximately an order of magnitude larger cross sections—than the $S=1$ $K^0$ component. Importantly, $\overline{K}^0$ scattering provides direct access to hyperons. In particular, the $K_L$-induced reaction $\overline{K}^0p\to \pi^+\Sigma^0$ is isospin-selective, proceeding purely via the $I=1$ channel, in contrast to the $K^- p\to \pi^\pm\Sigma^\mp$ reactions, which involve both $I=0$ and $I=1$ components. This makes it a valuable tool for isolating and studying $I = 1$ $\Sigma^*$ states. 

Using effective Lagrangian approach, we analyze the purely isospin $I=1$ reaction $K_Lp\to \pi^+\Sigma^0$. 
We perform a fit to the available experimental data on differential cross sections and $\Sigma^0$ recoil polarization, despite the limited precision and consistency of the data. Our model includes the contributions from well-established four-star resonances: $\Sigma(1189) 1/2^-$, $\Sigma(1385) 3/2^+$, $\Sigma(1670)3/2^-$, $\Sigma(1775)5/2^-$, along with $u$-channel nucleon exchange and $t$-channel $K^*$ exchange. The inclusion of the established four-star $\Sigma$ resonances is found to be necessary for describing the data. 

Furthermore, we investigated the contributions of several less-established $\Sigma$ resonances. 
Our analysis yields an optimal fit ($\chi^2/\mathrm{D.o.F}=1.606$) when including the three-star $\Sigma(1660) 1/2^+$, the one-star $\Sigma(1580)3/2^-$, and a $\Sigma^*(1/2^-)$ state. 

The results provide complementary support for the existence of $\Sigma(1660)1/2^+$ with fitted $m$ ($\approx$1696 MeV) and $\Gamma$ ($\approx$108 MeV) consistent with previous studies~\cite{Gao:2010ve, Gao:2012zh} and PDG values. 
Significantly, unlike previous analyses primarily focused on the $\pi\Lambda$ final state, our study of the $\pi\Sigma$ channel indicates the necessity of a $\Sigma^*(1/2^-)$ resonance with a mass around 1.54 GeV (fitted  $m\approx$1541 MeV, $\Gamma\approx$129 MeV) to adequately describe the $K_Lp\to \pi^+\Sigma^0$ data, which is compatible with Ref.~\cite{Zhang:2013sva, Kamano:2015hxa}. 
This finding strongly suggests the presence of the $\Sigma(1620) 1/2^-$ or a similar state contributing significantly in this energy region and channel. 
The inclusion of $\Sigma(1580) 3/2^-$ also improves the fit, with parameters consistent with PDG information. 
However, the current data show weak sensitivity to the $\Sigma(1750)3/2^-$ state. The model's prediction for the total cross section also shows good agreement with experimental data, further validating our approach. 

Due to the significant uncertainties and inconsistencies in the current historical data, precise determination of all resonance parameters remains challenging. Future high-precision measurements of $K_Lp$ scattering, such as by the KLF Collaboration at JLab using the GlueX spectrometer, potentially combined with analyses of related channels like $K_Lp\to \pi^+\Sigma^0$ and $K_Lp\to \pi^+\Lambda$, are crucial for clarifying the $\Sigma$ resonance spectrum, particularly confirming the properties and existence of the $\Sigma(1620) 1/2^-$ state. 

In addition, isolating the isospin-1 $T^1(\overline{K}N\to\pi\Sigma)$ amplitude from $K_L$ induced ractions allows for the extraction of isospin-0 $T^0(\overline{K}N\to\pi\Sigma)$ amplitude from the $K^- p\to \pi^\pm\Sigma^\mp$ processes, which benefit from more abundant data. 
This decomposition can further aid in the analysis of $\Lambda^*$ resonance properties.

\section{Acknowledgement}
D.~Guo thanks Xiong-Hui Cao for valuable discussions.
D.~Guo is supported by the National Natural Science Foundation of China under Grants No. 12347119, and the China Postdoctoral Science Foundation under Grant No. 2023M740117, and the China National Postdoctoral Program for Innovative Talent No. BX20230379. 
J. Shi is supported by the Natural Science Foundation of China under Grant No.
12105108 and the Guangdong Major Project for Basic and Applied Basic Research under Grant No. 2020B0301030008.
This work was supported in part by the U.~S.~Department of Energy, Office of Science, Office of Nuclear Physics, under Award No.~DE--SC0016583.

\section*{Data availability}

The experimental data that support the findings of this article are openly available in Refs.~\cite{Cho:1975dv, Engler:1978rr, Burkhardt:1975sd, Bologna-Edinburgh-Glasgow-Pisa-Rutherford:1977cyz, Engler:1976cn, Yamartino:1974sm}.

\end{document}